\documentclass{article}
\usepackage{amsmath}
\usepackage{amsfonts}
\usepackage{amssymb}
\renewcommand{\theequation}{\thesection.\arabic{equation}}
\makeatletter
\@addtoreset{equation}{section}
\makeatother
\makeatletter
\newlength{\earraycolsep}
\def\eqnarray{\stepcounter{equation}\let\@currentlabel\theequation
\global\@eqnswtrue\m@th
\global\@eqcnt\z@\tabskip\@centering\let\\\@eqncr
$$\halign to\displaywidth\bgroup\@eqnsel\hskip\@centering
$\displaystyle\tabskip\z@{##}$&\global\@eqcnt\@ne
\hskip 2\earraycolsep \hfil$\displaystyle{##}$\hfil
&\global\@eqcnt\tw@ \hskip 2\earraycolsep
$\displaystyle\tabskip\z@{##}$\hfil
\tabskip\@centering&\llap{##}\tabskip\z@\cr}
\makeatother

\begin{document}

\title{Integrable discretizations of the sine-Gordon equation}
\author{M. Boiti, F. Pempinelli, B. Prinari\\Dipartimento di Fisica dell'Universit\`{a} and Sezione INFN,\\73100 Lecce, Italy
\and A. Spire\\Physique Math\'{e}matique et Th\'{e}orique, CNRS-UMR5825,\\Universit\'{e} Montpellier 2, 34095 Montpellier, France}
\date{March 10, 2002}
\maketitle
\begin{abstract}
The inverse scattering theory for the sine-Gordon equation discretized in
space and both in space and time is considered.
\end{abstract}

%
%
%
%

\section{Introduction}

In the framework of integrable nonlinear evolution equations, Hirota recovered
almost 25 years ago the Bianchi superposition formula for the solutions of the
sine-Gordon equation, showing that it can be considered as an integrable
doubly discrete (both in space and time) version of the sine-Gordon equation
\cite{Hirota}. He found also its Lax pair, B\"{a}cklund transformations and
$N$-soliton solutions. In the following, this study was generalized to include
the discrete (in space or time) case \cite{Orfanidis}, soliton solutions were
obtained by using the dressing method \cite{Ragnisco}, the related inverse
spectral transform was studied \cite{Levi-Pilloni} and the nature of the
numerical instabilities of the solutions was discussed
\cite{AblowitzHerbst,AblowitzHerbst2}.

Here we are interested in the spectral transform for the discrete and doubly
discrete sine-Gordon equation with solutions decaying to $0$
($\operatorname{mod}\pi$) at space infinity .

The aim of this paper is twofold.

From one side, we re-examine the scattering theory of the spectral problem
introduced by Orfanidis in \cite{Orfanidis} which depends on a discretized
space variable $n$. According to the usual scheme, we introduce two summation
equations which define a couple of Jost solutions $\mu_{n}$ and $\nu_{n}$
characterized by their asymptotic behaviour, respectively, at $n\rightarrow
-\infty$ and $n\rightarrow+\infty$. These summation equations are, as usual,
used to define the spectral data. However, we show that, in contrast with the
continuous case, in order to study the analytical properties of the Jost
solution $\nu_{n}$ it is necessary to introduce an additional summation
equation that only in the continuous limit reduces to the previous one. Then,
we find the conditions on the potential that ensure the existence and
analyticity properties of the Jost solutions, we formulate the inverse problem
as a Riemann-Hilbert boundary value problem on the real axis and we derive the
time evolution of the spectral data, paying special attention to the nature of
the singularity of the time evolution of the solutions at $t=0$.

From the other side, we show that both in the semi-discrete and
doubly-discrete case, the equations of the matrix Orfanidis spectral problem
can be decoupled and reduced to a scalar problem, giving as principal spectral
problem the ``exact discretization'' of the Schr\"{o}dinger operator
introduced by Shabat \cite{Shabat}%
\begin{equation}
\varphi_{n+2}=g_{n}\varphi_{n+1}-(1+\lambda)\varphi_{n} \label{Shabat}%
\end{equation}
and obtained iterating the Darboux transformations. Then, following a
procedure analogous to that used in \cite{2+1sineGordon} for getting a 2+1
generalization of the sine-Gordon equations, we do not consider a Lax pair but
a Lax triplet, i.e. we derive the discrete sine-Gordon equations as a
compatibility condition of (\ref{Shabat}) with a pair of auxiliary spectral problems.

The spectral theory for the operator (\ref{Shabat}) was exhaustively studied
in \cite{BPPS} for a potential $g_{n}$, not necessarily real, satisfying
\begin{equation}
\sum_{n=-\infty}^{+\infty}(1+|n|)|g_{n-1}-2|<\infty. \label{gcondition}%
\end{equation}
and applied to a discrete version of the KdV. However, one must be advised
that the solutions of the discrete sine-Gordon equations are related to the
potential $g_{n}$ via an equation that can be considered a discretized version
of a Riccati equation and can be solved in terms of a continous fraction.
Moreover, the potential $g_{n}$ is complex and the problem of characterizing
the spectral data is left open.

By using this theory we introduce the spectral data, we find their time
evolution and, therefore, according to the usual inverse scattering scheme,
the Cauchy initial value problem is linearized. In particular, we show that
the time evolution is discontinuous at the initial time $t=0$.

We are also able to find the couple of doubly discrete auxiliary spectral
operators that in triplet with the doubly discrete Schr\"{o}dinger operator
gives the Hirota-Bianchi fourth order doubly discrete sine-Gordon equation.

The whole theory can be trivially extended to the sinh-Gordon case. Then, the
potential $g_{n}$ is real and the characterization equations for the spectral
data are easily obtained in analogy with the continuous case.

\section{Orfanidis Lax pair}

Let us consider the Lax pair proposed by Orfanidis \cite{Orfanidis}
\begin{align}
\chi_{n+1,m}  &  =M_{n,m}\chi_{n,m}\label{1}\\
\chi_{n,m+1}  &  =N_{n,m}\chi_{n,m} \label{2}%
\end{align}
with
\begin{align}
M_{n,m}  &  =\left(
\begin{array}
[c]{cc}%
e^{-i\left(  \theta_{n+1,m}-\theta_{n,m}\right)  /2} & ik\\
ik & \,e^{i\left(  \theta_{n+1,m}-\theta_{n,m}\right)  /2}%
\end{array}
\right) \label{Lax1}\\
N_{nm}  &  =\left(
\begin{array}
[c]{cc}%
1\, & \frac{\gamma}{ik}\,e^{-i\left(  \theta_{n,m+1}+\theta_{n,m}\right)
/2}\\
\frac{\gamma}{ik}\,e^{i\left(  \theta_{n,m+1}+\theta_{n,m}\right)  /2} & 1
\end{array}
\right)  \label{Lax22}%
\end{align}
with $\theta_{n,m}$ real, $k$ the spectral parameter and $\gamma$ a real
constant. The compatibility condition
\[
F\chi_{n+1,m}=E\chi_{n,m+1}
\]
where $E$ shifts $n$ and $F$ shifts $m$ gives by inserting in it (\ref{1}) and
(\ref{2}) $M_{n,m+1}N_{n,m}=N_{n+1,m}M_{n,m}$. From it we have the
Bianchi-Hirota equation
\begin{equation}
\sin\left(  \frac{\theta_{n+1,m+1}-\theta_{n+1,m}-\theta_{n,m+1}+\theta_{n,m}
}{4}\right)  =\gamma\sin\left(  \frac{\theta_{n+1,m+1}+\theta_{n+1,m}
+\theta_{n,m+1}+\theta_{n,m}}{4}\right)  .
\end{equation}
If we introduce
\begin{equation}
\theta_{i,j}=\theta_{i}(t+\tau j),\quad\quad t=m\tau,
\end{equation}
change the constant $\gamma$ as follows
\begin{equation}
\gamma\rightarrow\tau\gamma
\end{equation}
and take the limit $\tau\rightarrow0$ and $m\rightarrow\infty$ at $t$ fixed we
get the semi-discrete sine-Gordon equation in the light cone coordinates
\begin{equation}
\partial_{t}\theta_{n+1}-\partial_{t}\theta_{n}=4\gamma\sin\frac{1}{2}\left(
\theta_{n+1}+\theta_{n}\right)  . \label{semisine}%
\end{equation}
Notice that this equation can be obtained as compatibility condition of the
Lax pair
\begin{align}
\chi_{n+1}  &  =M_{n}\chi_{n}\label{semispace}\\
\partial_{t}\chi_{n}  &  =N_{n}\chi_{n} \label{semitime}%
\end{align}
with
\begin{align}
M_{n}  &  =\left(
\begin{array}
[c]{cc}%
e^{-i\left(  \theta_{n+1}-\theta_{n}\right)  /2} & ik\\
ik & \,e^{i\left(  \theta_{n+1}-\theta_{n}\right)  /2}%
\end{array}
\right) \\
N_{n}  &  =\left(
\begin{array}
[c]{cc}%
0\, & \frac{\gamma}{ik}\,e^{-i\theta_{n}}\\
\frac{\gamma}{ik}\,e^{i\theta_{n}} & 0
\end{array}
\right)
\end{align}

Notice also that in both cases, doubly discrete and space discrete cases, the
principal spectral problem (\ref{1}) is the same.

\section{Direct scattering problem}

\subsection{Jost solutions and summation equations}

Let us rewrite the spectral problem (\ref{1}) as
\begin{equation}
\chi_{n+1}-(\mathbf{1}+ik\sigma_{1})\chi_{n}=Q_{n}\chi_{n} \label{1next}%
\end{equation}
where the Pauli $\sigma$ matrices are defines as usual, i.e.
\begin{equation}
\sigma_{1}=\left(
\begin{array}
[c]{cc}%
0 & 1\\
1 & 0
\end{array}
\right)  ,\quad\sigma_{2}=\left(
\begin{array}
[c]{cc}%
0 & -i\\
i & 0
\end{array}
\right)  ,\quad\sigma_{3}=\left(
\begin{array}
[c]{cc}%
1 & 0\\
0 & -1
\end{array}
\right)
\end{equation}
and
\begin{equation}
Q_{n}=e^{-\frac{i}{2}\sigma_{3}(\theta_{n+1}-\theta_{n})}-\mathbf{1}%
\end{equation}
and let us consider it for $k=k_{\operatorname{Re}}$.

If $G_{n}$ is a Green function, i.e. if $G_{n}$ satisfies
\begin{equation}
G_{n+1}-(1+ik\sigma_{1})G_{n}=\delta_{n,0}\mathbf{1} \label{Gneq}%
\end{equation}
a matrix solution of (\ref{1next}) is given by the solution of the following
summation equation
\begin{equation}
\chi_{n}=w_{n}+\sum_{j=-\infty}^{+\infty}G_{n-j}Q_{j}\chi_{j}%
\end{equation}
where $w_{n}$ is a solution of the homogeneous equation
\begin{equation}
w_{n+1}-(\mathbf{1}+ik\sigma_{1})w_{n}=0.
\end{equation}
We represent $G_{n}$ and $\delta_{n,0}$ as Fourier integrals
\begin{align}
G_{n}  &  =\frac{1}{2\pi i}\oint_{|p|=R}p^{n-1}\widehat{G}(p)dp,\quad
\label{GnFourier}\\
\delta_{n,0}  &  =\frac{1}{2\pi i}\oint_{|p|=R}p^{n-1}dp. \label{deltaFourier}%
\end{align}
Inserting into (\ref{Gneq}) we get
\begin{equation}
((p-1)\mathbf{1}-ik\sigma_{1})\widehat{G}(p)=\mathbf{1}%
\end{equation}
and therefore
\begin{equation}
\widehat{G}(p)=\frac{1}{(p-1)^{2}+k^{2}}((p-1)\mathbf{1}+ik\sigma_{1}).
\end{equation}
Notice that $\widehat{G}(p)$ has a pole at $p=1\pm ik$ and notice also that at
large $p$
\begin{equation}
\left(  \mathbf{1}-\frac{\mathbf{1}+ik\sigma_{1}}{p}\right)  ^{-1}=\sum
_{j=0}^{+\infty}\frac{\left(  \mathbf{1}+ik\sigma_{1}\right)  ^{j}}{p^{j}}%
\end{equation}
and therefore at large $p$
\begin{equation}
\widehat{G}(p)=\sum_{j=0}^{+\infty}\frac{\left(  \mathbf{1}+ik\sigma
_{1}\right)  ^{j}}{p^{j+1}}. \label{Gp}%
\end{equation}
We consider $k=k_{\operatorname{Re}}$ and for different choices of $R$, i.e.
for $R>1$ and $R<1$, we get different Green functions, $G$ and $H$, and,
correspondingly, different summation equations defining different Jost
solutions, $\phi$ and $\psi$.

For $R>1+k^{2}$ the integrand in (\ref{GnFourier}) inside the disk $|p|\leq R$
contains a pole at $p=0$ for $n\leq0$ and the poles $p=1\pm ik$ and it is
analytic outside the disk. The integral can be computed evaluating the
residuum at $p=\infty$ by using (\ref{Gp}) and we get
\begin{equation}
G_{n}=\Theta(n-1)\left(  \mathbf{1}+ik\sigma_{1}\right)  ^{n-1}%
\end{equation}
where $\Theta(n)$ is the discrete version of the Heaviside function%
\begin{equation}
\Theta(n)=\left\{
\begin{array}
[c]{c}%
1\qquad n\geq0\\
0\qquad n<0
\end{array}
\right.  .
\end{equation}
Therefore
\begin{equation}
\phi_{n}=w_{n}+\sum_{j=-\infty}^{n-1}\left(  \mathbf{1}+ik\sigma_{1}\right)
^{n-j-1}Q_{j}\phi_{j}.
\end{equation}
If $R<1$ we have to subtract to the previous integral the residua at the poles
$p=1\pm ik$. We have
\begin{equation}
H_{n}=\Theta(n-1)\left(  \mathbf{1}+ik\sigma_{1}\right)  ^{n-1}-\frac{1}{2}
(\mathbf{1}+\sigma_{1})(1+ik)^{n-1}-\frac{1}{2}(\mathbf{1}-\sigma
_{1})(1-ik)^{n-1}.
\end{equation}
Since
\begin{equation}
\left(  \mathbf{1}+ik\sigma_{1}\right)  ^{n}=\frac{1}{2}(\mathbf{1}+\sigma
_{1}) (1+ik)^{n}+\frac{1}{2}(\mathbf{1}-\sigma_{1})(1-ik)^{n} \label{identity}%
\end{equation}
we have
\begin{equation}
H_{n}=-\Theta(-n)\left(  \mathbf{1}+ik\sigma_{1}\right)  ^{n-1}%
\end{equation}
and the summation equation becomes%
\begin{equation}
\psi_{n}=w_{n}-\sum_{j=n}^{+\infty}\left(  \mathbf{1}+ik\sigma_{1}\right)
^{n-j-1}Q_{j}\psi_{j}. \label{psi}%
\end{equation}
It is convenient to choose
\begin{equation}
w_{n}=(\mathbf{1}-i\sigma_{2})(\mathbf{1}+ik\sigma_{3})^{n} \label{wn}%
\end{equation}
so that $\phi_{n}$ and $\psi_{n}$ satisfy the following boundary conditions
\begin{align}
\phi_{n}  &  \sim\left(  \mathbf{1}-i\sigma_{2}\right)  \left(  \mathbf{1}
+ik\sigma_{3}\right)  ^{n}\qquad\qquad n\rightarrow-\infty\label{phi-phibar}\\
\psi_{n}  &  \sim\left(  \mathbf{1}-i\sigma_{2}\right)  \left(  \mathbf{1}
+ik\sigma_{3}\right)  ^{n}\qquad\qquad n\rightarrow+\infty. \label{psi-psibar}%
\end{align}
In order to study the analytical properties of the Jost solutions with respect
to the spectral parameter $k$, it is necessary to introduce the modified
matrix Jost solutions
\begin{align}
\mu_{n}  &  =\phi_{n}\left(  \mathbf{1}+ik\sigma_{3}\right)  ^{-n}%
\label{phimu}\\
\nu_{n}  &  =\psi_{n}\left(  \mathbf{1}+ik\sigma_{3}\right)  ^{-n}
\label{psinu}%
\end{align}
and, then, to consider separately the two columns of these matrices, which we
denote as follows
\begin{equation}
\mu_{n}=(\mu_{n}^{-},\mu_{n}^{+}),\quad\nu_{n}=(\nu_{n}^{+},\nu_{n}^{-}).
\end{equation}
They satisfy the difference equations
\begin{align}
\left(  1\mp ik\right)  \mu_{n+1}^{\pm}  &  =\left[  \mathbf{1}+ik\sigma
_{1}+Q_{n}\right]  \mu_{n}^{\pm}\label{diff1}\\
\left(  1\pm ik\right)  \nu_{n+1}^{\pm}  &  =\left[  \mathbf{1}+ik\sigma
_{1}+Q_{n}\right]  \nu_{n}^{\pm} \label{diff2}%
\end{align}
and have constant asymptotic behaviour
\begin{align}
\mu_{n}  &  \sim\left(  \mathbf{1}-i\sigma_{2}\right)  \qquad\qquad
n\rightarrow-\infty\label{asM}\\
\nu_{n}  &  \sim\left(  \mathbf{1}-i\sigma_{2}\right)  \qquad\qquad
n\rightarrow+\infty. \label{asN}%
\end{align}

\subsection{Existence and analyticity of the Jost solutions $\mu^{\pm}$}

We will show that the Jost solutions $\mu_{n}^{\pm}(k)$ defined by the
summation equations
\begin{align}
\mu_{n}^{+}(k)  &  =\left(
\begin{array}
[c]{c}%
-1\\
1
\end{array}
\right)  +\sum_{j=-\infty}^{n-1}\frac{\left(  \mathbf{1}+ik\sigma_{1}\right)
^{n-j-1}}{(1-ik)^{n-j}}Q_{j}\mu_{j}^{+}(k)\label{Mint}\\
\mu_{n}^{-}(k)  &  =\left(
\begin{array}
[c]{c}%
1\\
1
\end{array}
\right)  +\sum_{j=-\infty}^{n-1}\frac{\left(  \mathbf{1}+ik\sigma_{1}\right)
^{n-j-1}}{(1+ik)^{n-j}}Q_{j}\mu_{j}^{-}(k) \label{Nint}%
\end{align}
are analytic, correspondingly, in the upper half plane and in the lower half
plane of the spectral parameter $k$ and continuous for $k_{\operatorname{Im}
}\geq0$ and $k_{\text{Im}}\leq0$.

Let us write the solution of the integral equation (\ref{Mint}) in the form of
a Neumann series
\begin{equation}
\mu_{n}^{+}(k)=\sum_{\ell=0}^{+\infty}C_{n}^{\ell}(k)\qquad C_{n}^{\ell
}(k)=\left(
\begin{array}
[c]{c}%
C_{n}^{\ell,(1)}(k)\\
C_{n}^{\ell,(2)}(k)
\end{array}
\right)  \label{Neumann}%
\end{equation}
where
\begin{equation}
C_{n}^{0}(k)=\left(
\begin{array}
[c]{c}%
-1\\
1
\end{array}
\right)  ,\qquad C_{n}^{\ell+1}(k)=\sum_{j=-\infty}^{n-1}\frac{\left(
\mathbf{1}+ik\sigma_{1}\right)  ^{n-j-1}}{(1-ik)^{n-j}}Q_{j}C_{j}^{\ell}(k)
\end{equation}
or, in component form, taking into account the identity (\ref{identity})
\begin{align}
C_{n}^{\ell+1,(1)}(k)  &  =\frac{1}{2(1-ik)}\sum_{j=-\infty}^{n-1}\left\{
\left[  1+\left(  \frac{1+ik}{1-ik}\right)  ^{n-j-1}\right]  \left(
e^{-\frac{i}{2}\left(  \theta_{j+1}-\theta_{j}\right)  }-1\right)  C_{j}
^{\ell,(1)}(k)\right. \nonumber\\
&  -\left.  \left[  1-\left(  \frac{1+ik}{1-ik}\right)  ^{n-j-1}\right]
\left(  e^{\frac{i}{2}\left(  \theta_{j+1}-\theta_{j}\right)  }-1\right)
C_{j}^{\ell,(2)}(k)\right\} \\
C_{n}^{\ell+1,(2)}(k)  &  =\frac{1}{2(1-ik)}\sum_{j=-\infty}^{n-1}\left\{
-\left[  1-\left(  \frac{1+ik}{1-ik}\right)  ^{n-j-1}\right]  \left(
e^{-\frac{i}{2}\left(  \theta_{j+1}-\theta_{j}\right)  }-1\right)  C_{j}
^{\ell,(1)}(k)\right. \nonumber\\
&  +\left.  \left[  1+\left(  \frac{1+ik}{1-ik}\right)  ^{n-j-1}\right]
\left(  e^{\frac{i}{2}\left(  \theta_{j+1}-\theta_{j}\right)  }-1\right)
C_{j}^{\ell,(2)}(k)\right\}  .
\end{align}
The series (\ref{Neumann}) is formally a solution of the discrete integral
equation (\ref{Mint}). One can prove by induction on $\ell\in\mathbb{N}$ that
for $k_{\operatorname{Im}}\geq0$
\begin{equation}
\left|  C_{n}^{\ell,(j)}(k)\right|  \leq\frac{2^{\ell}}{\left|  1-ik\right|
^{\ell}}\frac{1}{\ell!}\left[  \sum_{l=-\infty}^{n-1}\left|  e^{\frac{i}
{2}(\theta_{l+1}-\theta_{l})}-1\right|  \right]  ^{\ell},\quad\quad j=1,2.
\label{bound}%
\end{equation}
Indeed, for each component $j=1,2$ we have
\begin{align*}
\left|  C_{n}^{\ell+1,(j)}(k)\right|   &  \leq\frac{1}{2\left|  1-ik\right|
}\sum_{l=-\infty}^{n-1}\left|  e^{-\frac{i}{2}\left(  \theta_{l+1}-\theta
_{l}\right)  }-1\right|  \left[  \left|  1+\left(  \frac{1+ik}{1-ik}\right)
^{n-l-1}\right|  \left|  C_{l}^{\ell,(1)}(k)\right|  \right. \\
&  +\left.  \left|  1-\left(  \frac{1+ik}{1-ik}\right)  ^{n-l-1}\right|
\left|  C_{l}^{\ell,(2)}(k)\right|  \right]
\end{align*}
where we used that the potential $\theta_{n}$ is real. Moreover
\begin{equation}
\left|  1\pm\left(  \frac{1+ik}{1-ik}\right)  ^{n-l-1}\right|  \leq1+\left|
\frac{1+ik}{1-ik}\right|  ^{n-l-1}\leq2
\end{equation}
since $k_{\operatorname{Im}}>0$ and $n-l-1\geq0$. Therefore we have
\begin{equation}
\left|  C_{n}^{\ell+1,(j)}(k)\right|  \leq\frac{1}{\left|  1-ik\right|  }
\sum_{l=-\infty}^{n-1}\left|  e^{-\frac{i}{2}\left(  \theta_{l+1}-\theta
_{l}\right)  }-1\right|  \left[  \left|  C_{l}^{\ell,(1)}(k)\right|  +\left|
C_{l}^{\ell,(2)}(k)\right|  \right]
\end{equation}
and we can use the inductive hypothesis to get
\begin{equation}
\left|  C_{n}^{\ell+1,(j)}(k)\right|  \leq\frac{1}{\ell!}\frac{2^{\ell+1}
}{\left|  1-ik\right|  ^{\ell+1}}\sum_{l=-\infty}^{n-1}\left|  e^{-\frac{i}
{2}\left(  \theta_{l+1}-\theta_{l}\right)  }-1\right|  \left(  \sum
_{l_{1}=-\infty}^{\ell-1}\left|  e^{-\frac{i}{2}\left(  \theta_{l_{1}
+1}-\theta_{l_{1}}\right)  }-1\right|  \right)  ^{\ell}. \label{ind+1}%
\end{equation}
The use of the summation by parts formula, according to which for any real
non-negative sequence $\left\{  b_{j}\right\}  _{j=-\infty}^{+\infty}$ such
that the series $\sum_{j=-\infty}^{+\infty}b_{j}$ is convergent and for any
$m\in\mathbb{N}_{0}$
\begin{equation}
\sum_{k=-\infty}^{n}b_{k}\left(  \sum_{j=-\infty}^{k-1}b_{j}\right)  ^{m}
\leq\frac{1}{m+1}\left(  \sum_{j=-\infty}^{n-1}b_{j}\right)  ^{m+1},
\end{equation}
completes the proof of (\ref{bound}).

Therefore for $k_{\operatorname{Im}}\geq0$
\begin{equation}
\left|  C_{n}^{\ell,(j)}(k)\right|  \leq\frac{P\left(  n\right)  ^{\ell}}
{\ell!},\qquad j=1,2
\end{equation}
where
\begin{equation}
P(n)=2\sum_{j=-\infty}^{n-1}\left|  e^{\frac{i}{2}(\theta_{j+1}-\theta_{j}
)}-1\right|
\end{equation}
and we conclude that the Neumann series (\ref{Neumann}) for $\mu_{n}^{+}(k)$
is uniformly convergent (with respect to $n$ and $k$ in the upper half-plane)
and
\begin{equation}
\left|  \mu_{n}^{+,(j)}(k)\right|  \leq e^{P(\infty)}\qquad j=1,2
\end{equation}
provided the potential $\theta_{n}$ is such that
\begin{equation}
P(\infty)=2\sum_{j=-\infty}^{+\infty}\left|  e^{\frac{i}{2}(\theta
_{j+1}-\theta_{j})}-1\right|  <+\infty. \label{Pcondition}%
\end{equation}
In the same way one can prove that $\mu_{n}^{-}$ is analytic for
$k_{\operatorname{Im}}<0$ and continuous for $k_{\operatorname{Im}}\leq0$.

\subsection{Existence and analyticity of the Jost solutions $\nu^{\pm}$}

Let us now consider the summation equations defining the Jost solutions
$\nu^{\pm}$, which can be rewritten as%

\begin{align}
\nu_{n}^{+}(k)  &  =\left(
\begin{array}
[c]{c}%
1\\
1
\end{array}
\right)  -\frac{1}{1+ik}\sum_{j=n}^{+\infty}\frac{\left(  \mathbf{1}
-ik\sigma_{1}\right)  ^{j-n+1}}{(1-ik)^{j-n+1}}Q_{j}\nu_{j}^{+}
(k)\label{nu+first}\\
\nu_{n}^{-}(k)  &  =\left(
\begin{array}
[c]{c}%
-1\\
1
\end{array}
\right)  -\frac{1}{1-ik}\sum_{j=n}^{+\infty}\frac{\left(  \mathbf{1}
-ik\sigma_{1}\right)  ^{j-n+1}}{(1+ik)^{j-n+1}}Q_{j}\nu_{j}^{-} (k)
\label{nu-first}%
\end{align}
showing explicitly that their kernel is singular at $k=\pm i$, i.e. in both
the lower and upper half planes of $k$. This dissymmetry of the summation
equations for the Jost solutions $\nu^{\pm}$ with respect to $\mu^{\pm}$ is
peculiar of the discrete case, since it disappears in the continuous limit. In
fact, in order to study the analytic properties of the Jost solutions one
needs to define the same solutions by introducing alternative summation
equations, which can be obtained by exploiting the special structure of the
potential $Q_{n}$ and the symmetry property of the spectral problem
(\ref{1next}) for $n\rightarrow-n$.

By using the relation
\begin{equation}
Q_{n}+\sigma_{1}Q_{n}\sigma_{1}+\sigma_{1}Q_{n}\sigma_{1}Q_{n} =0
\label{QQrelation}%
\end{equation}
one can easily verify that
\begin{equation}
(\mathbf{1}+ik\sigma_{1}+Q_{n})^{-1}=\frac{1}{1+k^{2}}(\mathbf{1}-ik\sigma
_{1}+\sigma_{1}Q_{n}\sigma_{1}) \label{inverse}%
\end{equation}
and therefore the spectral problem (\ref{1next}) can be rewritten as%
\begin{equation}
\frac{1}{1+k^{2}}(\mathbf{1}-ik\sigma_{1}+\sigma_{1}Q_{n}\sigma_{1})\chi
_{n+1}=\chi_{n}.
\end{equation}
If we introduce
\begin{equation}
\chi_{n+1}=(1+k^{2})^{n}\xi_{-n}%
\end{equation}
we have
\begin{equation}
\xi_{n+1}=(\mathbf{1}-ik\sigma_{1}+\sigma_{1}Q_{-n}\sigma_{1})\xi_{n}%
\end{equation}
and, consequently, using the same procedure we followed above for $\phi_{n}$
we get
\begin{equation}
\xi_{n}=\xi_{0n}+\sum_{j=-\infty}^{n-1}\left(  \mathbf{1}-ik\sigma_{1}\right)
^{n-j-1}\sigma_{1}Q_{-j}\sigma_{1}\xi_{j}%
\end{equation}
and coming back to $\chi_{n}$
\begin{equation}
\chi_{n}=w_{n}+\sum_{\ell=n+1}^{+\infty}(1+k^{2})^{n-\ell}(\mathbf{1}
-ik\sigma_{1})^{\ell-n-1}\sigma_{1}Q_{\ell-1}\sigma_{1}\chi_{\ell} \label{chi}%
\end{equation}
where we choose $w_{n}$ as in (\ref{wn}).

One can check explicitly that $\psi_{n}$ defined in (\ref{psi}) satisfies this
summation equation by moving in (\ref{psi}) the $j=n$ term of the sum from the
right side to the left side, by applying from the left $(\mathbf{1}
+ik\sigma_{1}+Q_{n})^{-1}(\mathbf{1}+ik\sigma_{1})$ and then by using on the
right (\ref{inverse}), (\ref{QQrelation}) and the spectral equation for
$\psi_{n}$.

Therefore from (\ref{chi}) by considering the transformation (\ref{psinu}) we
get the following alternative summation equations for $\nu^{\pm}(k)$%

\begin{align}
\nu_{n}^{+}(k)  &  =\left(
\begin{array}
[c]{c}%
1\\
1
\end{array}
\right)  +\sum_{j=n+1}^{+\infty}\frac{\left(  \mathbf{1}-ik\sigma_{1}\right)
^{j-n-1}}{(1-ik)^{j-n}}\sigma_{1}Q_{j-1}\sigma_{1}\nu_{j}^{+}(k)\\
\nu_{n}^{-}(k)  &  =\left(
\begin{array}
[c]{c}%
-1\\
1
\end{array}
\right)  +\sum_{j=n+1}^{+\infty}\frac{\left(  \mathbf{1}-ik\sigma_{1}\right)
^{j-n-1}}{(1+ik)^{j-n}}\sigma_{1}Q_{j-1}\sigma_{1}\nu_{j}^{-}(k).
\end{align}

Following a procedure analogous to that one used for the $\mu^{\pm}(k)$ one
can prove that the $\nu^{\pm}(k)$ are also analytic, correspondingly, in the
upper half plane and in the lower half plane of the spectral parameter $k$ and
continuous for $k_{\operatorname{Im}}\geq0$ and $k_{\text{Im}}\leq0$ provided
the potential satisfies the condition (\ref{Pcondition}).

Finally, let us note that, taking into account the explicit expression of the
Green's functions, one can easily obtain the asymptotic behavior of the Jost
solutions at large $k$
\begin{align}
\mu_{n}^{+}(k)  &  \sim\left(
\begin{array}
[c]{c}%
-1\\
1
\end{array}
\right)  \qquad\nu_{n}^{+}(k)\sim\left(
\begin{array}
[c]{c}%
1\\
1
\end{array}
\right)  \qquad\left|  k\right|  \rightarrow\infty\text{\qquad}
k_{\operatorname{Im}}>0\label{ask}\\
\mu_{n}^{-}(k)  &  \sim\left(
\begin{array}
[c]{c}%
1\\
1
\end{array}
\right)  \qquad\nu_{n}^{-}(k)\sim\left(
\begin{array}
[c]{c}%
-1\\
1
\end{array}
\right)  \qquad\left|  k\right|  \rightarrow\infty\text{\qquad}%
k_{\operatorname{Im}}<0. \label{ask2}%
\end{align}

\subsection{Scattering data}

Let us define the Wronskian of any two vectors $\ v$ and $w$ as
\begin{equation}
W\left(  v,w\right)  =\det\left(  v,w\right)  .
\end{equation}
The vector-valued sequences $v_{n}$ and $w_{n}$ are linearly independent if
$W\left(  v_{n},w_{n}\right)  \neq0$ for all $n$.

In particular, if $v_{n}$ and $w_{n}$ are any two solutions of the scattering
problem (\ref{1next}), their Wronskian satisfies the recursive relation
\begin{equation}
W\left(  v_{n+1},w_{n+1}\right)  =\left(  1+k^{2}\right)  W\left(  v_{n}
,w_{n}\right)  .
\end{equation}
Hence, for any positive integer $j$
\begin{align}
W\left(  \phi_{n}^{+}(k),\phi_{n}^{-}(k)\right)   &  =\left(  1+k^{2}\right)
^{j}W\left(  \phi_{n-j}^{+}(k),\phi_{n-j}^{-}(k)\right) \nonumber\\
&  =\left(  1+k^{2}\right)  ^{n}W\left(  \mu_{n-j}^{+}(k),\mu_{n-j}
^{-}(k)\right)
\end{align}
and
\begin{align}
W\left(  \psi_{n}^{+}(k),\psi_{n}^{-}(k)\right)   &  =\left(  1+k^{2}\right)
^{-j}W\left(  \psi_{n+j}^{+}(k),\psi_{n+j}^{-}(k)\right) \nonumber\\
&  =\left(  1+k^{2}\right)  ^{n}W\left(  \nu_{n+j}^{+}(k),\nu_{n+j}
^{-}(k)\right)  .
\end{align}
Taking into account (\ref{asM}) and (\ref{asN}), for $k=k_{\operatorname{Re}}$
in the limit $j\rightarrow\infty$ we get
\begin{equation}
W\left(  \phi_{n}^{+}(k),\phi_{n}^{-}(k)\right)  =-2\left(  1+k^{2}\right)
^{n},\qquad W\left(  \psi_{n}^{+}(k),\psi_{n}^{-}(k)\right)  =2\left(
1+k^{2}\right)  ^{n} \label{W3}%
\end{equation}
which shows that the eigenfunctions $\phi_{n}^{+}$ and $\phi_{n}^{-}$ are
linearly independent, as well as $\psi_{n}^{+}$ and $\psi_{n}^{-}$. Since the
discrete scattering problem (\ref{1next}) is a second-order difference
equation, there are at most two linearly independent solutions for any fixed
value of $k$ and consequently we can write $\phi_{n}^{+}$ and $\phi_{n}^{-}$
as linear combination of $\psi_{n}^{+}$ and $\psi_{n}^{-}$ or vice-versa. The
coefficients of this linear combinations depend on $k$. Hence the relations
($k=k_{\operatorname{Re}}$)
\begin{equation}
\phi_{n}^{\pm}(k)=b^{\pm}(k)\psi_{n}^{\pm}(k)+a^{\pm}(k)\psi_{n}^{\mp}(k)
\label{data2}%
\end{equation}
which define the scattering data $a^{\pm}(k)$, $b^{\pm}(k)$.

In terms of the Jost solutions, they can also be written as
\begin{equation}
\frac{\mu_{n}^{\pm}(k)}{a^{\pm}(k)}=\rho^{\pm}(k)\left(  \frac{1\pm ik}{1\mp
ik}\right)  ^{n}\nu_{n}^{\pm}(k)+\nu_{n}^{\mp}(k) \label{data4}%
\end{equation}
where we introduced the reflection coefficients
\begin{equation}
\rho^{\pm}(k)=\frac{b^{\pm}(k)}{a^{\pm}(k)}. \label{refelction}%
\end{equation}
Notice that from (\ref{data4}), by using (\ref{asN}), we can get the
asymptotics of $\mu_{n}^{\pm}$ for $n\rightarrow+\infty$
\begin{equation}
\mu_{n}^{\pm}(k)\sim b^{\pm}(k)\left(  \frac{1\pm ik}{1\mp ik}\right)
^{n}\binom{\pm1}{1}+a^{\pm}(k)\binom{\mp1}{1}. \label{asM+}%
\end{equation}

Calculating $W\left(  \phi_{n}^{+}(k),\phi_{n}^{-}(k)\right)  $ using
(\ref{data2}) results in
\begin{equation}
W\left(  \phi_{n}^{+}(k),\phi_{n}^{-}(k)\right)  =-\left(  a^{+}
(k)a^{-}(k)-b^{+}(k)b^{-}(k)\right)  W\left(  \psi_{n}^{+}(k),\psi_{n}
^{-}(k)\right)
\end{equation}
and taking into account (\ref{W3}) yields
\begin{equation}
a^{+}(k)a^{-}(k)-b^{+}(k)b^{-}(k)=1\qquad k\in\mathbb{R}. \label{char1}%
\end{equation}
Moreover,
\begin{equation}
W\left(  \phi_{n}^{\pm}(k),\psi_{n}^{\pm}(k)\right)  =\mp2\left(
1+k^{2}\right)  ^{n}a^{\pm}(k) \label{W2}%
\end{equation}
or, equivalently,
\begin{equation}
a^{\pm}(k)=\mp\frac{1}{2}W\left(  \mu_{n}^{\pm}(k),\nu_{n}^{\pm}(k)\right)
\end{equation}
which proves, from one side, that $a^{+}(k)$ can be analytically continued in
the upper half plane (respectively $a^{-}(k)$ can be analytically continued in
the lower half plane) and, from the other side, that the zeros of $a^{+}(k)$
in the upper half plane (respectively of $a^{-}(k)$) correspond to bound
states of the scattering problem.

We note that for real potentials $\theta_{n}$, if $\chi_{n}(k)$ satisfies the
scattering problem (\ref{1next}), then
\begin{equation}
\tilde{\chi}_{n}(k)=i\sigma_{2}\chi_{n}^{\ast}(k^{\ast})
\end{equation}
satisfies the same equation. Taking into account the boundary conditions
(\ref{phi-phibar}) and (\ref{psi-psibar}), we conclude that if the potential
$\theta_{n}$ is real the Jost solutions obey the symmetry conditions
\begin{equation}
\phi_{n}^{-}(k)=i\sigma_{2}\left(  \phi_{n}^{+}(k^{\ast})\right)  ^{\ast
}\qquad\psi_{n}^{-}(k)=-i\sigma_{2}\left(  \psi_{n}^{+}(k^{\ast})\right)
^{\ast}. \label{symm1}%
\end{equation}
Moreover, one can easily show by using (\ref{W2}) that
\begin{equation}
a^{-}(k)=\left(  a^{+}(k^{\ast})\right)  ^{\ast} \label{symm02}%
\end{equation}
and by using (\ref{data2}) for $k=k_{\operatorname{Re}}$ that
\begin{equation}
\qquad b^{-}(k)=-\left(  b^{+}(k)\right)  ^{\ast}. \label{symm2}%
\end{equation}

The scattering coefficients can also be given as explicit sum of the Jost
solutions. The formulae are derived as follows. First, we rewrite for
$k=k_{\operatorname{Re}}$ the summation equation (\ref{Mint}) as
\begin{align}
\mu_{n}^{+}(k)  &  =\left(
\begin{array}
[c]{c}%
-1\\
1
\end{array}
\right)  +\sum_{j=-\infty}^{+\infty}\frac{\left(  1+ik\sigma_{1}\right)
^{n-j-1}}{(1-ik)^{n-j}}Q_{j}\mu_{j}^{+}(k)\nonumber\\
&  -\frac{1}{(1-ik)}\sum_{j=n}^{+\infty}\frac{\left(  1-ik\sigma_{1}\right)
^{j-n+1}}{(1+ik)^{j-n+1}}Q_{j}\mu_{j}^{+}(k)
\end{align}
and, then, using (\ref{nu-first}) we obtain
\begin{align}
\mu_{n}^{+}(k)-\nu_{n}^{-}(k)a^{+}(k)  &  =\left(
\begin{array}
[c]{c}%
-1\\
1
\end{array}
\right)  \left(  1-a^{+}(k)\right)  +\sum_{j=-\infty}^{+\infty}\frac{\left(
1+ik\sigma_{1}\right)  ^{n-j-1}}{(1-ik)^{n-j}}Q_{j}\mu_{j}^{+}(k)\nonumber\\
&  -\frac{1}{(1-ik)}\sum_{j=n}^{+\infty}\frac{\left(  1-ik\sigma_{1}\right)
^{j-n+1}}{(1+ik)^{j-n+1}}Q_{j}\left[  \mu_{j}^{+}(k)-\nu_{j}^{-}
(k)a^{+}(k)\right]  .
\end{align}
Inserting in it (\ref{data4}) we have
\begin{align}
&  b^{+}(k)\left(  \frac{1+ik}{1-ik}\right)  ^{n}\left[  \nu_{n}^{+}
(k)-\frac{1}{(1+ik)}\sum_{j=n}^{+\infty}\frac{\left(  1-ik\sigma_{1}\right)
^{j-n+1}}{(1-ik)^{j-n+1}}Q_{j}\nu_{j}^{+}(k)\right] \nonumber\\
&  \quad\quad=\left(
\begin{array}
[c]{c}%
-1\\
1
\end{array}
\right)  \left(  1-a^{+}(k)\right)  +\sum_{j=-\infty}^{+\infty}\frac{\left(
1+ik\sigma_{1}\right)  ^{n-j-1}}{(1-ik)^{n-j}}Q_{j}\mu_{j}^{+}(k).
\end{align}
From (\ref{nu+first}) it follows that the term in square brackets in the
left-hand side is $\left(  1\ ,1\right)  ^{T}$ and therefore we obtain using
(\ref{identity})
\begin{align}
a^{+}(k)  &  =1+\frac{1}{2(1-ik)}\sum_{j=-\infty}^{+\infty}\left[  -\left(
e^{-\frac{i}{2}\left(  \theta_{j+1}-\theta_{j}\right)  }-1\right)  \mu
_{j}^{+,(1)}(k)\right. \nonumber\\
&  \quad\quad+\left.  \left(  e^{\frac{i}{2}\left(  \theta_{j+1}-\theta
_{j}\right)  }-1\right)  \mu_{j}^{+,(2)}(k)\right] \label{a}\\
b^{+}(k)  &  =\frac{1}{2(1-ik)}\sum_{j=-\infty}^{+\infty}\left(  \frac
{1+ik}{1-ik}\right)  ^{-j-1}\left[  \left(  e^{-\frac{i}{2}\left(
\theta_{j+1}-\theta_{j}\right)  }-1\right)  \mu_{j}^{+,(1)}(k)\right.
\nonumber\\
&  \quad\quad+\left.  \left(  e^{\frac{i}{2}\left(  \theta_{j+1}-\theta
_{j}\right)  }-1\right)  \mu_{j}^{+,(2)}(k)\right]  . \label{b}%
\end{align}
These (discrete version of) integral representations, together with the
symmetry relations (\ref{symm02}) and (\ref{symm2}) completely determine the
scattering coefficients.

The discrete scattering problem (\ref{1next}) can possess discrete eigenvalues
(bound states). These occur whenever $a^{+}(k_{j}^{+})=0$ for some $k_{j}^{+}$
in the upper half plane or $a^{-}(k_{\ell}^{-})=0$ for some $k_{\ell}^{-}$ in
the lower half plane. Indeed, for such values of the spectral parameter from
(\ref{W2}) it follows that $W\left(  \phi_{n}^{+}(k_{j}^{+}),\psi_{n}
^{+}(k_{j}^{+})\right)  =0$ and $W\left(  \phi_{n}^{-}(k_{\ell}^{-}),\psi
_{n}^{-}(k_{\ell}^{-})\right)  =0$ and therefore the eigenfunctions are
linearly dependent, i.e.
\begin{align}
\phi_{n}^{+}(k_{j}^{+})  &  =b_{j}^{+}\psi_{n}^{+}(k_{j}^{+})\qquad
j=1,\dots,J^{+}\label{prop}\\
\phi_{n}^{-}(k_{\ell}^{-})  &  =b_{\ell}^{-}\psi_{n}^{-}(k_{\ell}^{-}
)\qquad\ell=1,\dots,J^{-} \label{propp}%
\end{align}
for some complex constants $b_{j}^{\pm}$. In terms of the modified Jost
solutions (\ref{prop}), (\ref{propp}) can be written as
\begin{equation}
\mu_{n}^{+}(k_{j}^{+})=b_{j}^{+}\left(  \frac{1+ik_{j}^{+}}{1-ik_{j}^{+}
}\right)  ^{n}\nu_{n}^{+}(k_{j}^{+}),\qquad\mu_{n}^{-}(k_{\ell}^{-})=b_{\ell
}^{-}\left(  \frac{1-ik_{\ell}^{-}}{1+ik_{\ell}^{-}}\right)  ^{n}\nu_{n}
^{-}(k_{\ell}^{-}). \label{prop2}%
\end{equation}
Note that when the potential $\theta_{n}$ is real, (\ref{symm02}) implies that
if $k_{j}^{+}$ is a zero of $a^{+}(k)$ in the upper half-plane, then
$k_{j}^{-}=\left(  k_{j}^{+}\right)  ^{\ast}$ is a zero of $a^{-}(k)$ in the
lower half-plane and therefore $J^{+}=J^{-}$.

\section{Inverse Scattering problem}

Let us assume $a^{+}(k)$ has $J^{+}$ simple zeros $\left\{  k_{j}^{+}\right\}
_{j=1}^{J^{+}}$ in the upper half plane and $a^{-}(k)$ has $J^{-}$ simple
zeros $\left\{  k_{j}^{-}\right\}  _{j=1}^{J^{-}}$ in the lower half plane.
Then, by (\ref{prop2}) it follows
\begin{align}
\text{\textrm{Res}}\left(  \frac{\mu_{n}^{+}(k)}{a^{+}(k)};k_{j}^{+}\right)
&  =\mu_{n}^{+}(k_{j}^{+})\left(  \left.  \frac{da^{+}}{dk}\right|
_{k=k_{j}^{+}}\right)  ^{-1}=C_{j}^{+}\left(  \frac{1+ik_{j}^{+}}{1-ik_{j}
^{+}}\right)  ^{n}\nu_{n}^{+}(k_{j}^{+})\label{Res1}\\
\text{\textrm{Res}}\left(  \frac{\mu_{n}^{-}(k)}{a^{-}(k)};k_{j}^{-}\right)
&  =\mu_{n}^{-}(k_{j}^{-})\left(  \left.  \frac{da^{-}}{dk}\right|
_{k=k_{j}^{-}}\right)  ^{-1}=C_{j}^{-}\left(  \frac{1-ik_{j}^{-}}{1+ik_{j}
^{-}}\right)  ^{n}\nu_{n}^{-}(k_{j}^{-}) \label{Res2}%
\end{align}
where
\begin{equation}
C_{j}^{+}=b_{j}^{+}\left(  \left.  \frac{da^{+}}{dk}\right|  _{k=k_{j}^{+}
}\right)  ^{-1},\qquad C_{j}^{-}=b_{j}^{-}\left(  \left.  \frac{da^{-}}
{dk}\right|  _{k=k_{j}^{-}}\right)  ^{-1}. \label{norming}%
\end{equation}
Taking into account the analytic properties of the Jost solutions and of the
scattering coefficients $a^{\pm}(k)$, as well as the asymptotic behavior
(\ref{ask}), we can use the Cauchy-Green formula to obtain from the ``jump''
conditions (\ref{data4})
\begin{align}
\nu_{n}^{-}(k)  &  =\left(
\begin{array}
[c]{c}%
-1\\
1
\end{array}
\right)  +\sum_{j=1}^{J^{+}}\left(  \frac{1+ik_{j}^{+}}{1-ik_{j}^{+}}\right)
^{n}\frac{C_{j}^{+}}{k-k_{j}^{+}}\nu_{n}^{+}(k_{j}^{+})\nonumber\\
&  +\frac{1}{2\pi i}\int_{-\infty}^{+\infty}\frac{\rho^{+}(\zeta)}{\zeta
-k+i0}\left(  \frac{1+i\zeta}{1-i\zeta}\right)  ^{n}\nu_{n}^{+}(\zeta
)d\zeta\label{RH1}\\
\nu_{n}^{+}(k)  &  =\left(
\begin{array}
[c]{c}%
1\\
1
\end{array}
\right)  +\sum_{j=1}^{J^{-}}\left(  \frac{1-ik_{j}^{-}}{1+ik_{j}^{-}}\right)
^{n}\frac{C_{j}^{-}}{k-k_{j}^{-}}\nu_{n}^{-}(k_{j}^{-})\nonumber\\
&  -\frac{1}{2\pi i}\int_{-\infty}^{+\infty}\frac{\rho^{-}(\zeta)}{\zeta
-k-i0}\left(  \frac{1-i\zeta}{1+i\zeta}\right)  ^{n}\nu_{n}^{-}(\zeta)d\zeta.
\label{RH2}%
\end{align}
Equations (\ref{RH1}) and (\ref{RH2}) define a Riemann-Hilbert problem on the
line which, in principle, allows one to solve the inverse problem, i.e. to
reconstruct the Jost solutions from the scattering data.

In order to complete the inverse problem, we need to reconstruct the potential
from the scattering data. Let us rewrite the difference equations
(\ref{diff2}) in matrix form
\begin{equation}
\left[  1+ik\sigma_{3}\right]  \nu_{n+1}=\left[  1+ik\sigma_{1}+Q_{n}\right]
\nu_{n} \label{diff3}%
\end{equation}
and let us write the asymptotic expansion at large $k$ of $\nu_{n}=\left(
\nu_{n}^{+},\nu_{n}^{-}\right)  $
\begin{equation}
\nu_{n}(k)=\nu_{n}^{(0)}+\frac{\nu_{n}^{(-1)}}{ik}+\dots\label{exp}%
\end{equation}
where, strictly speaking, the expansion of the first column is performed for
$k_{\operatorname{Im}}>0$ while the expansion of the second one is performed
for $k_{\operatorname{Im}}<0$. Inserting this expansion into (\ref{diff3}) and
taking into account that, according to (\ref{ask}) and (\ref{ask2}),
\ $\nu_{n}^{(0)}=1-i\sigma_{2}$ yields
\begin{equation}
Q_{n}=\frac{1}{2}\left[  \nu_{n+1}^{(-1)}\sigma_{3}-\sigma_{1}\nu_{n}
^{(-1)}\right]  \left(  1+i\sigma_{2}\right)  . \label{invpot}%
\end{equation}

\section{Time evolution of spectral data}

The time evolution of the Jost solutions and, consequently, of the scattering
data is fixed by (\ref{semitime}).

Since the matrix $\mu_{n}(1+i\sigma_{3}k)^{n}$ is a general solution of
(\ref{semispace}) there exists a $\Omega(t;k)$ such that
\begin{equation}
\chi_{n}(t;k)=\mu_{n}(t;k)(1+i\sigma_{3}k)^{n}\Omega(t;k),
\label{timesolution}%
\end{equation}
satisfies both (\ref{semispace}) and (\ref{semitime}).

Inserting (\ref{timesolution}) into (\ref{semitime}), if $\theta
_{n}\rightarrow r\pi$ ($r\in\mathbb{N}$) for $n\rightarrow\pm\infty$, by
evaluating the limit $n\rightarrow-\infty$ we get, recalling (\ref{asM}),
\begin{equation}
\Omega(t;k)=\exp\left(  -i\sigma_{3}\frac{\eta\gamma}{k}t\right)
\end{equation}
where $\eta=(-1)^{r}$.

Then, we evaluate the limit for $n\rightarrow+\infty$ by using (\ref{asM+})
and we obtain the time evolution of spectral data ($k=k_{\operatorname{Re}}$)
\begin{equation}
b_{t}^{\pm}(k)=\mp\frac{2i\eta\gamma}{k\mp i0\eta\gamma t}b^{\pm}(k),\qquad
a_{t}^{\pm}(k)=0
\end{equation}
i.e.
\begin{equation}
b^{\pm}(k,t)=b^{\pm}(k,0)e^{\mp2i\frac{\eta\gamma}{k}t},\qquad a^{\pm
}(k,t)=a^{\pm}(k,0).
\end{equation}
For the discrete spectral data one gets that the locations of poles at
$k=k_{j}^{\pm}$ are fixed while
\begin{equation}
C_{j}^{\pm}(t)=C_{j}^{\pm}(0)e^{\mp2i\frac{\eta\gamma}{k_{j}^{\pm}}t}.
\end{equation}

Notice that the switching on of the time evolution introduces a singularity at
$k=0$. One can prove that this does not spoil the good properties of the
spectral transform. In the following section where the matrix spectral problem
is reduced to a scalar problem this will be shown explicitly.

\section{Lax triplet}

\subsection{Semi-discrete case}

We can write the spectral problem (\ref{1next}) in component-form and decouple
the two equations to get
\begin{align}
\chi_{n+2}^{(1)}  &  =g_{n}\chi_{n+1}^{(1)}-(1+k^{2})\chi_{n}^{(1)}%
\label{discrSchr}\\
\chi_{n}^{(2)}  &  =\frac{1}{ik}\left[  \chi_{n+1}^{(1)}-e^{-\frac{i}{2}
(\theta_{n+1}-\theta_{n})}\chi_{n}^{(1)}\right]  \label{2comp}%
\end{align}
with
\begin{equation}
g_{n}=e^{\frac{i}{2}(\theta_{n+1}-\theta_{n})}+e^{-\frac{i}{2}(\theta
_{n+2}-\theta_{n+1})} \label{gn}%
\end{equation}
i.e. we recover the discrete Schr\"{o}dinger equation whose scattering theory
was given in \cite{BPPS}. In this case the time evolution is fixed by the pair
of spectral problems
\begin{align}
k^{2}\partial_{t}\chi_{n}^{(1)}  &  =-\gamma e^{-i\theta_{n}}\chi_{n+1}
^{(1)}+\gamma e^{-\frac{i}{2}\left(  \theta_{n+1}+\theta_{n}\right)  }\chi
_{n}^{(1)}\label{Lax4}\\
\partial_{t}\chi_{n+1}^{(1)}  &  =e^{-\frac{i}{2}\left(  \theta_{n+1}
-\theta_{n}\right)  }\partial_{t}\chi_{n}^{(1)}+\gamma e^{-i\theta_{n+1}}
\chi_{n}^{(1)}. \label{Lax3}%
\end{align}
The compatibility requirement for the three spectral problems (\ref{discrSchr}
), (\ref{Lax4}) and (\ref{Lax3}) furnishes after one integration the
semi-discrete sine-Gordon equation (\ref{semisine}).

Note that, if we introduce
\begin{equation}
\alpha_{n}=e^{-\frac{i}{2}(\theta_{n+1}-\theta_{n})},
\end{equation}
the equation defining $g_{n}$ can be rewritten as
\begin{equation}
g_{n}=\alpha_{n+1}+\frac{1}{\alpha_{n}}, \label{gtilde}%
\end{equation}
that can be considered as a discrete Riccati equation, which can be solved for
$\alpha_{n}$ in terms of a continuous fraction of $g_{n}$ (see \cite{Wall}).

Note also that for $\theta_{n}$ vanishing at large $n$ ($\operatorname{mod}
\pi$) we have from (\ref{gn}) that $g_{n}-2\rightarrow0$ and the spectral
theory developed in \cite{BPPS} is applicable. However, the condition required
to be satisfied by $g_{n}-2$ in \cite{BPPS} is stricter than (\ref{Pcondition}).

We conclude that the Orfanidis matrix spectral problem can be reduced to a
scalar Schr\"{o}dinger spectral problem if the potential $\theta
_{n}\rightarrow0$ ($\operatorname{mod}\pi$) sufficiently rapidly.

\subsection{Time evolution of spectral data}

The evolution of the spectral data can be determined by considering the first
and the third Lax operators, i.e. the spectral problem (\ref{discrSchr}) and
(\ref{Lax4}).

In order to meet the notations used in \cite{BPPS} we introduce the
eigenfunction as follows
\begin{equation}
\chi_{n}^{(1)}(k)=(1+ik)^{n}\xi_{n}(k). \label{psi-mu}%
\end{equation}
The principal spectral problem (\ref{discrSchr}) reads
\begin{equation}
(1+ik)\xi_{n+2}+(1-ik)\xi_{n}=g_{n}\xi_{n+1} \label{mod-eq}%
\end{equation}
while the auxiliary spectral problem (\ref{Lax4}) becomes
\begin{equation}
k^{2}\partial_{t}\xi_{n}=\gamma e^{-\frac{i}{2}(\theta_{n+1}+\theta_{n})}%
\xi_{n}-(1+ik)\gamma e^{-i\theta_{n}}\xi_{n+1}. \label{chit}%
\end{equation}
We need the Jost solution $\mu_{n}^{+}(k)$ introduced in \cite{BPPS}, which is
analytic in the upper half-plane and is defined via the following discrete
integral equation
\begin{equation}
\mu_{n}^{+}(k)=1-\frac{1}{2ik}\sum_{j=n+1}^{+\infty}\left[  1-\left(
\frac{1+ik}{1-ik}\right)  ^{j-n}\right]  (g_{j-1}-2)\mu_{j}^{+}(k)
\label{int-mu}%
\end{equation}
and which has the following asymptotic behaviour for $n\rightarrow+\infty$
\begin{equation}
\lim_{n\rightarrow+\infty}\mu_{n}^{+}(k)=1,\quad k_{\operatorname{Im}}\geq0
\label{mu+}%
\end{equation}
and for $n\rightarrow-\infty$
\begin{equation}
\mu_{n}^{+}(k)\sim a^{+}(k)+\left(  \frac{1-ik}{1+ik}\right)  ^{n}%
b^{+}(k),\qquad k_{\operatorname{Im}}=0 \label{nu+}%
\end{equation}
where $a^{+}(k)$ is the inverse of the transmission coefficient and $b^{+}(k)$
is the reflection coefficient.

Then we look for a solution of (\ref{chit}) of the form
\begin{equation}
\xi_{n}(t;k)=\Omega(t;k)\mu_{n}^{+}(t;k) \label{Cmu}%
\end{equation}
If we substitute (\ref{Cmu}) into (\ref{chit}), we get
\begin{equation}
k^{2}\left[  \Omega_{t}\Omega^{-1}\mu_{n}^{+}+\partial_{t}\mu_{n}^{+}\right]
=\gamma e^{-\frac{i}{2}(\theta_{n+1}+\theta_{n})}\mu_{n}^{+}-(1+ik)\gamma
e^{-i\theta_{n}}\mu_{n+1}^{+}. \label{Ct}%
\end{equation}
If $\theta_{n}\rightarrow r\pi$ ($r\in\mathbb{N}$) for $n\rightarrow\pm\infty
$, taking into account the asymptotic behaviour of $\mu_{n}^{+}$ as
$n\rightarrow+\infty$ we get, first,
\begin{equation}
\Omega_{t}\Omega^{-1}=-\frac{i\eta\gamma}{k+i0\eta\gamma t},\quad\eta=(-1)^{r}%
\end{equation}
and, then, by considering the limit for $n\rightarrow-\infty$
($k_{\operatorname{Im}}=0$) the evolution equation for the spectral data
\begin{align}
a_{t}^{+}(k)  &  =0\\
b_{t}^{+}(k)  &  =\frac{2i\eta\gamma}{k-i0\eta\gamma t}b^{+}(k),
\end{align}
which can be trivially integrated to
\begin{align}
a(t;k)  &  =a(0;k)\\
b(t;k)  &  =b(0;k)\exp\left[  \frac{2i\eta\gamma}{k}t\right]  . \label{bt}%
\end{align}

Let us now study in details the behaviour of the Jost solutions at $k=0$ when
the time is switched on. For the sake of simplicity, let us take $\eta
\gamma=1$, i.e. let us scale the time.

In \cite{BPPS} the Riemann-Hilbert boundary value problem defining the Jost
solutions was given as
\begin{align}
\mu_{n}^{-}(t;k)  &  =1+\frac{1}{2\pi i}\int_{-\infty}^{+\infty}\left(
\frac{1-is}{1+is}\right)  ^{n}\frac{\mu_{n}^{-}(t;-s)\rho^{+}(t,s)}%
{s-k+i0}ds\label{riem1}\\
\frac{\mu_{n}^{+}(t;k)}{a^{+}(k)}  &  =1+\frac{1}{2\pi i}\int_{-\infty
}^{+\infty}\left(  \frac{1-is}{1+is}\right)  ^{n}\frac{\mu_{n}^{-}%
(t;-s)\rho^{+}(t,s)}{s-k-i0}ds. \label{riem2}%
\end{align}
For the sake of simplicity, we omit a possible contribution from the discrete
part of the spectrum. It can be added without difficulty along the same lines
followed for the Orfanidis spectral problem.

For convenience we introduce
\begin{align}
\Phi_{n}(k)  &  =\Omega^{-1}(k)\partial_{t}\xi_{n}^{-}(k)=\Omega
^{-1}(k)\left[  \Omega(k)\mu_{n}^{-}(k)\right]  _{t}\nonumber\\
&  =-\frac{i}{k+i0t}\mu_{n}^{-}(k)+\partial_{t}\mu_{n}^{-}(k). \label{Phi}%
\end{align}
Deriving (\ref{riem1}) with respect to time we have
\begin{align}
\partial_{t}\mu_{n}^{-}(k)  &  =\frac{1}{2\pi i}\int_{-\infty}^{+\infty
}\left(  \frac{1-is}{1+is}\right)  ^{n}\frac{\mu_{n}^{-}(-s)\rho^{+}%
(s)}{s-k+i0} \frac{2i}{s-i0t}ds\nonumber\\
&  +\int_{-\infty}^{+\infty}\left(  \frac{1-is}{1+is}\right)  ^{n}
\frac{\left(  \partial_{t}\mu_{n}^{-}(-s)\right)  \rho^{+}(s)}{s-k+i0}ds
\end{align}
and therefore
\begin{align}
\Phi_{n}(k)  &  =-\frac{i}{k+i0t}+\frac{1}{2\pi i}\int_{-\infty}^{+\infty
}\left(  \frac{1-is}{1+is}\right)  ^{n}\frac{\mu_{n}^{-}(-s)\rho^{+}
(s)}{s-k+i0}\left(  \frac{i}{s-i0t}-\frac{i}{k+i0t}\right)  ds\nonumber\\
&  +\frac{1}{2\pi i}\int_{-\infty}^{+\infty}\left(  \frac{1-is}{1+is}\right)
^{n}\frac{\Phi_{n}(-s)\rho^{+}(s)}{s-k+i0}.
\end{align}
Since
\begin{equation}
\frac{1}{s-i0t}-\frac{1}{k+i0t}=\frac{k-s}{\left(  k+i0t\right)  \left(
s-i0t\right)  }%
\end{equation}
we have finally
\begin{equation}
\Phi_{n}(k)=-\frac{1}{k+i0t}S_{n}(t)+\frac{1}{2\pi i}\int_{-\infty}^{+\infty
}\left(  \frac{1-is}{1+is}\right)  ^{n}\frac{\Phi_{n}(-s)\rho^{+}(s)} {s-k+i0}
\label{intPhi}%
\end{equation}
where
\begin{equation}
S_{n}(t)=1+\frac{1}{2\pi i}\int_{-\infty}^{+\infty}\left(  \frac{1-is}
{1+is}\right)  ^{n}\frac{\mu_{n}^{-}(-s)\rho^{+}(s)}{s-i0t}=\left\{
\begin{array}
[c]{c}%
\frac{\mu_{n}^{+}(0)}{a^{+}(0)}\qquad\text{for }t>0\\
\mu_{n}^{-}(0)\qquad\text{for }t<0
\end{array}
\right.  .
\end{equation}
In order to study the singularity of $\Phi_{n}(k)$ at $k=0$ one can try to
solve the integral equation by iteration and see at each step how the
singularity $1/\left(  k-i0t\right)  $ is transformed.

Let us first consider the case $t>0$. Iterating once yields an integral
equation that contains the distribution
\begin{equation}
\frac{1}{s-i0}\frac{1}{s-k+i0}=\frac{1}{k-i0}\left(  \frac{1}{s-k+i0}-\frac
{1}{s-i0}\right)  .
\end{equation}
Therefore in this case the iteration renormalizes the coefficient of the
singularity $1/\left(  k-i0t\right)  $ but does not change the nature of the
singularity at $k=0$.

Let us now consider the case $t<0$. In this case in iterating we get in the
integral
\begin{equation}
\frac{1}{s+i0}\frac{1}{s-k+i0}%
\end{equation}
which is a distribution continuous at $k=0.$ Therefore in this case we can
conclude that the singularity of $\Phi_{n}(k)$ can be singled out and it is
just
\begin{equation}
-\frac{i}{k-i0t}\mu_{n}^{-}(0).
\end{equation}
Recalling the definition of $\Phi_{n}$ we conclude that, for $t<0$,
$\partial_{t}\mu_{n}^{-}(k)$, thanks to the continuity of $\mu_{n}^{-}(k)$ at
$k=0$, is less singular than $1/k$.

In conclusion, $k\partial_{t}\mu_{n}^{-}(k)$ and, therefore, $k^{2}
\partial_{t}\xi_{n}^{-}(k)$ appearing in the auxiliary spectral problem, are
continuous at $k=0$. Analogously of course for $k\partial_{t}\mu_{n}^{+}(k)$
and $k^{2}\partial_{t}\xi_{n}^{+}(k)$. They are, however, discontinuous at
$t=0$, i.e. $g_{n}(t)$ and, consequently, $\theta_{n}(t)$ evolve according to
different laws for $t\lessgtr0$, which is not surprising since the sine-Gordon
equation is not an evolution equation.

\subsection{Doubly discrete case}

Analogously in the system of the two spectral problems (\ref{1}) and (\ref{2})
the two components $\chi_{n,m}^{(1)}$ and $\chi_{n,m}^{(2)}$ of $\chi_{n,m}$
can be decoupled. Of course $\chi_{n,m}^{(1)}$ satisfy the same discrete
version of the Schr\"{o}dinger spectral problem as $\chi_{n}^{(1)}$, i.e.
\begin{equation}
\chi_{n+2,m}^{(1)}=g_{n,m}\chi_{n+1,m}^{(1)}-(1+k^{2})\chi_{n,m}^{(1)}
\label{Lax00}%
\end{equation}
where
\begin{equation}
g_{n,m}=e^{\frac{i}{2}(\theta_{n+1,m}-\theta_{n,m})}+e^{-\frac{i}{2}
(\theta_{n+2,m}-\theta_{n+1,m})},
\end{equation}
while the time evolution is fixed by the couple of spectral problems
\begin{align}
k^{2}\chi_{n,m+1}^{(1)}  &  =\left(  k^{2}+\gamma e^{-\frac{i}{2}
(\theta_{n+1,m}+\theta_{n,m+1})}\right)  \chi_{n,m}^{(1)}-\gamma e^{-\frac
{i}{2}(\theta_{n,m+1}+\theta_{n,m})}\chi_{n+1,m}^{(1)}\label{Lax30}\\
\chi_{n+1,m+1}^{(1)}  &  =\chi_{n+1,m}^{(1)}+e^{-\frac{i}{2}(\theta
_{n+1,m+1}-\theta_{n,m+1})}\chi_{n,m+1}^{(1)}\nonumber\\
&  \quad\quad\quad+\left(  \gamma e^{\frac{i}{2}(\theta_{n,m+1}+\theta_{n,m}
)}-e^{-\frac{i}{2}(\theta_{n+1,m}-\theta_{n,m})}\right)  \chi_{n,m} ^{(1)}.
\label{Lax40}%
\end{align}

The compatibility among the three spectral problems (\ref{Lax00}),
(\ref{Lax30}) and (\ref{Lax40}) furnishes, after an integration, the
Hirota-Bianchi doubly discrete sine-Gordon equation.

\subsection{Time evolution of spectral data}

Following a procedure analogous to that followed in the semi-discrete case we
look for a solution of (\ref{Lax30}) of the form
\begin{equation}
\xi_{n,m}(k)=\Omega_{m}(k)\mu_{n,m}^{+}(k)(1+ik)^{n}%
\end{equation}
where $\mu_{n,m}^{+}(k)$ is the Jost solution introduced in \cite{BPPS}.
Taking into account the asymptotic behaviours (\ref{mu+}) and (\ref{nu+}), if
$\theta_{n,m}\rightarrow r\pi$ ($r\in\mathbb{N}$) for $n\rightarrow\pm\infty$,
we obtain, respectively, for $n\rightarrow+\infty$ the time evolution of
$\Omega$
\begin{equation}
\Omega_{m+1}(k)\Omega_{m}^{-1}(k)=1-\frac{i\eta\gamma}{k}%
\end{equation}
and for $n\rightarrow-\infty$ the time evolution of the spectral data
($k_{\operatorname{Im}}=0$)
\begin{align}
&  a_{m+1}(k)=a_{m}(k)\\
&  b_{m+1}(k)=-\frac{(\eta\gamma-ik)}{(\eta\gamma+ik)}b_{m}(k)
\end{align}
where $\eta$ is defined as above.

If we introduce
\begin{equation}
a_{m}=a(t+\tau m),\quad\quad b_{m}=b(t+\tau m),\quad\quad t=n\tau,
\end{equation}
change the constant $\gamma$ as follows
\begin{equation}
\gamma\rightarrow\tau\gamma
\end{equation}
and take the limit $\tau\rightarrow0$ and $n\rightarrow\infty$ at $t$ fixed we
recover, as expected, the time evolution of the spectral data for the
semi-discrete sine-Gordon equation.

\subsection{Acknowledgments}

This work was partially supported by PRIN 2000 ``Sintesi'' and was performed
in the framework of the INTAS project 99-1782. The authors acknowledge useful
critical remarks by an anonymous referee.

\end{document}